\documentclass[12pt,preprintnumbers,amsmath]{revtex4}
\usepackage{graphicx}
\usepackage{bm}
\begin{document}

\topmargin 0cm

\title{Open Inflationary Universes in Gauss-Bonnet Brane Cosmology}
\author{Sergio del Campo, Ram\'on Herrera and Joel Saavedra}
\address{Instituto de F\'{\i}sica, Pontificia Universidad
Cat\'olica de Valpara\'{\i}so, Casilla 4950, Valpara\'{\i}so.}
\date{\today}

\begin{abstract}
In this article, we study a type of one-field approach for open
inflationary universe scenario  in the context of  braneworld
models with a Gauss-Bonnet correction term.  For a one-bubble
universe model, we determine and characterize the existence of the
Coleman-De Lucia instanton together with the period of inflation
after tunneling has occurred. Our results are compared those
analogous obtained when the usual Einstein Theory of Gravitation
is used.

\end{abstract}
\maketitle

\section{\label{sec:level1} Introduction}

Recent observations from the WMAP~\cite{wmap} are entirely
consistent with a universe have a total energy density that is
very close to its critical value, where the total density
parameter has the value $\Omega =1.02\pm 0.04$. Most people
interpret this value as corresponding to a flat universe. But,
according to this result, we might take the alternative point of
view of have a marginally open or close universe
model~\cite{ellis-k}, which early presents an inflationary period
of expansion. At this point, we should mention that Ratra and
Peebles were the first to elaborate on the open inflation model
\cite{Ratra,Ratra2}. The approach of consider open or a closed
inlfationary universe models has already been considered in the
literature,~\cite{re6,re8,shiba,ramon,delCampo:2004gh, del
Campo:2004ee, Balart:2007je} in the context of the Einstein theory
of relativity, Jordan-Brans-Dicke (JBD) theory, Randall-Sundrum
type of gravitation theory, and tachyonic field cosmology
respectively. All these models have been worked out in a four
dimensional spacetime or on a four dimensional brane (where lived
matter contents) embedding in a five dimensional bulk. The idea of
considering an extra dimension has received great attention in the
last few years, since it is believed that these models shed light
on the solution to fundamental problems when the universe is
traced back to very early times. Specifically, the possibility of
creating an open or closed universe from the context of a Brane
World (BW) scenario \cite{MBPGD, MBPGD2, MBPGD3}, has quite
recently been considered.

BW cosmology offers a novel approach to our understanding of the
evolution of the universe. The most spectacular consequence of
this scenario is the modification of the Friedmann equation, in a
particular case when a five dimensional model is considered, and
where the matter described through a scalar field, is confined to
a four dimensional Brane, while gravity can be propagated in the
bulk. These kind of models can be obtained from superstring
theory~\cite{witten, witten2}. For a comprehensible review on BW
cosmology, see Refs.~\cite{lecturer,lecturer2,lecturer3} for
example. Specifically, consequences of a chaotic inflationary
universe scenario in a BW model was described~\cite{maartens},
where it was found that the slow-roll approximation is enhanced by
the modification of the Friedman equation.

 When a stage closed to the Big-Bang is studied, quantum effects
should be included in the bulk. In the high dimensional theory,
the high curvature correction terms should be added to the
Einstein Hilbert action. As far as it is well known,  the
correction term does not introduce ghost in this action  is the
Gauss-Bonnet (GB) combination. The GB term arises naturally as the
leading order of the $\alpha'$ expansion of heterotic superstring
theory, where, $\alpha'$ is the inverse string tension. The GB
term is given by

\begin{equation}
GB=R^{\mu \nu \lambda \rho }R_{^{\mu \nu \lambda \rho }}-4R^{\mu
\nu }R_{\nu \mu }+R^{2},  \label{q1}
\end{equation}
that often appears in higher dimensional version of gravity
theories, especially in string theory. In fact, all versions of
string theory (except Type II) in 10 dimensions (D = 10) include
this term. Therefore, could be interesting to make an study of the
effects of the Gauss-Bonnet term that produces on inflationary
braneworld models \cite{Meng:2003pn,Lidsey:2003sj,
Calcagni:2004as, Kim:2004gs, Kim:2004jc, Kim:2004hu,
Murray:2006fw}. On the other hand, Gauss-Bonnet term was consider
as an alternative for explain the mysterious of currently dark
energy observed\cite{Leith:2007bu, Koivisto:2006ai, Sanyal:2006wi,
Koivisto:2006xf, Carter:2005fu, Nojiri:2005am, Nojiri:2005vv}. The
purpose of the present paper is to study an open inflation
universe model, where the scalar field is confined to the four
dimensional Brane that is embedded in five dimensional bulk where
a Gauss-Bonnet (GB) contribution is considered.

The plan of the paper is as follows: In Sec. II we specify the
effective four dimensional cosmological equations from a five-Anti
de Sitter GB brane world  model. We write down the Euclidean field
equations, and then we solve them numerically. Here, the existence
of the Coleman-De Lucia (CDL) instanton is established. In Sec.
III we determine the characteristics of an open inflationary
universe model that is produced after tunneling process has
occurred. Our results are compared to those analogous results
obtained by using Einstein's theory of gravity. Finally, we
conclude in Sec. IV.


\section{\label{sec:level2}The Euclidean cosmological equations in
Randall-Sundrum type II Scenario}

We start with the action given by
\begin{eqnarray}
S &=&\frac{1}{2\kappa
_{5}^{2}}\int_{bulk}d^{5}x\sqrt{-g_{5}}\left\{ R-2\Lambda
_{5}+\alpha \left( R^{\mu \nu \lambda \rho }R_{^{\mu \nu \lambda
\rho }}-4R^{\mu \nu }R_{\nu \mu }+R^{2}\right) \right\} ,\nonumber \\
&&+\int_{brane}d^{4}x\sqrt{-g_{4}}\left(
\mathcal{L}_{matter}-\lambda \right),\label{action1}
\end{eqnarray}
where $\Lambda _{5}=-3\mu ^{2}\left( 2-4\alpha \mu ^{2}\right) $
is the cosmological constant in five dimensions, with the
$AdS_{5}$ energy scale $\mu=1/l$ and $\lambda$ corresponds to the
brane tension. $\mathcal{L}_{matter}$ is the matter lagrangian for
the inflaton field on the brane. $\kappa _{5}^{2}=8\pi/m_{5}$ is
the five dimensional gravitational coupling constant. Finally, the
GB coupling constant $\alpha$ can be related to the string scale
($g_{s}$) as $\alpha\simeq1/8g_{s}$ when the GB term is taken as
to the lowest-order stringy correction to the 5D gravity.
Cosmological dynamics on the brane arise due to its motion through
the static bulk space. If we consider a Friedmann-Robertson-Walker
brane, where the scale factor $a(t)$ describes the position of the
brane in the bulk.  The exact Friedmann-like equation is given by
\cite{davis}

\begin{equation}
2\mu \sqrt{1+\frac{H^{2}}{\mu ^{2}}}\left( 3-4\alpha \mu
^{2}+8\alpha H^{2}\right)+\frac{k}{a^{2}} =\kappa _{5}^{2}\left(
\rho +\lambda \right) ,  \label{q3}
\end{equation}
where $\rho$ represents the energy density of the matter sources
on the brane and $k=0,+1 and -1$ represents flat,closed and open
spatial section, respectively. The conservation equation of the
matter onto the brane follows directly from Gauss-Codazzi
equations. For a perfect fluid matter source, it is reduced to the
familiar form,
\begin{equation}
\dot{\rho}+3H\left( \rho +p\right) =0.  \label{q4}
\end{equation}
In term of the inflaton field Eq.(\ref{q4}) can be written as
\begin{equation}
\displaystyle \ddot{\phi}+3\frac{\dot{a}}{a}\dot{\phi}+V_{,\phi
}(\phi )=0\,\,, \label{ecphi}
\end{equation}
where $V(\phi)$ represents the effective potential associated to
the inflaton field. This potential is given by
\begin{equation}
V(\phi )=\frac{1}{2}m^{2}\left( 1+\frac{\alpha ^{2}}{\delta ^{2}+(\phi -v)^{2}%
}\right),   \label{q5}
\end{equation}
where $\alpha$, $\delta$ and $v$ are arbitrary constants, which
will be set by phenomenological considerations. This potential was
first considered by Linde\cite{re6} in order to obtain a
successful scenario for open inflation in Einstein's. The first
term of the effective potential controls inflation after quantum
tunnelling has occurred (its shape coincides with that used in the
simplest chaotic inflationary universe model, $m^{2}\phi
{^{2}}/2$), and the second term controls the bubble nucleation,
whose role is to create an appropriate shape in the inflaton
potential, $V(\phi)$ .i.e., the parameters $\alpha$, $\delta$ and
$v$ just affect the quantum tunelling process. Therefore, once
that tunelling occurs the inflationary dynamics is drives by the
simplest chaotic potential and the parameters of the second term
are not considering in the inflationary epoch. Following
Ref.~\cite{ramon} we take $\delta ^{2}=0.1$,
$\delta^{2}=2\alpha^{2}$, and $v = 3.5$, which, will provide the
adequate N e-folds number of inflation after the tunneling.
Certainly, this is not the only choice, since other values for
these parameters can also lead to a successful open inflationary
scenario (with any value of $\Omega $, in the range $0 < \Omega <
1$).

In particular, in our model the Gauss Bonnet Brane World the
inflaton field begins at the false vacuum ($\phi_{F}$) and then
decays to the value $\phi_{T}$ via quantum tunnelling through the
potential barrier (see Fig.1), generating during this process an
open inflationary universe. In order to study  the nucleation of
bubble we write down the corresponding Euclidean equations of
motion, and the $O(4)$ invariant Euclidean spacetime metric is
written as
\begin{equation}
\displaystyle d{s}^{2}\,=\,d{\tau }^{2}\,+\,a(\tau )^{2}\,(\,\,d{\psi }%
^{2}\,+\,\sin ^{2}\psi \,d{\Omega _{2}^{2}}\,\,).  \label{met}
\end{equation}
\begin{figure}[th]
\includegraphics[width=2.0in,angle=0,clip=true]{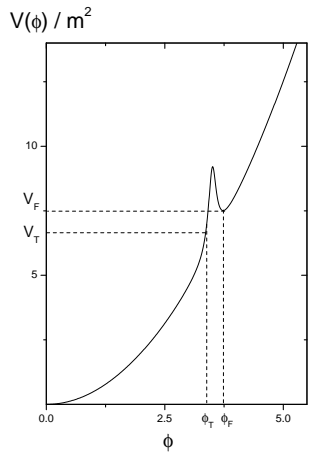}
\vspace{-1.0cm} \caption{The potential form for the inflaton
field, as in Ref. \cite{re6}, we will consider the theory with $m
= 1.5 \times 10^{-6}$, which is necessary to have a proper
amplitude for density perturbations during inflation in the model
with $q=1$. And $v = 3.5$, which, will provide the adequate N
e-folds number of inflation after the tunneling. By changing this
parameter by few percent one can get any value of $\Omega$ from 0
to 1. In particular we will take $\delta^{2}=2\alpha^{2}$ with
$\delta= 0.1$.\label{fig1}}
\end{figure}
Instead of using the complicated form for the exact Friedmann-like
equation (\ref{q3}), we used an effective Friedmann equation given
by
\begin{equation}
H^{2}-\frac{1}{a^{2}}=\beta _{q}^{2}\rho ^{q},  \label{friedmann2}
\end{equation}
where $\beta _{q}^{2}$ is a model dependent constant, whose values
are justified for the observation (in particular for the values of
N e-folds number after inflations and for the density of scalar
perturbations).
\begin{table}
\caption{\label{tab:table1}This table shows the values of $\beta$
in Eistein-Hilbert and Gauss Bonnet Brane world models. }
\begin{tabular}{||c||c||c||}
\hline\hline Model & $q$ & $\beta _{q}$ \\ \hline\hline Einstein
Hilbert & $1$ & $10$ \\ \hline\hline Gauss Bonnet Brane World &
$2/3$ & $10^{-3}$ \\ \hline\hline
\end{tabular}
\end{table}
Then, we can written the Euclidean field equations as follow,
\begin{eqnarray}
a^{\prime \prime } &=&-a\beta_{q} m^{2\left( q-1\right) }\left(
-\frac{\phi ^{\prime 2}}{2}+V(\phi )\right) ^{q}\left[
1+\frac{3}{2}q\frac{\phi ^{\prime 2}}{\left( -\frac{\phi ^{\prime
2}}{2}+V(\phi )\right) }\right] ,
\label{eqm1} \\
&&
\begin{array}{ll}
& \phi ^{\prime \prime }+3\frac{a^{\prime }}{a}\phi ^{\prime
}-V_{,\phi }=0,
\end{array}
\nonumber
\end{eqnarray}
where the upper prime represent derivative with respect to
Euclidean time ($\tau$) and $V_{,\phi }=\frac{dV}{d\phi}$.

\begin{figure}[th]
\includegraphics[width=3.0in,angle=0,clip=true]{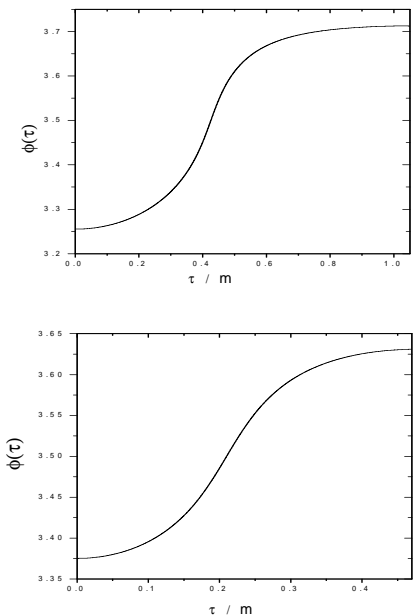}
\vspace{-1.7cm}
\caption{The instanton, $\phi(\tau)$, as a function of the Euclidean time $%
\tau $ is shown for Einstein GR and  BW. The upper panel shows the
case $\beta=10$, and the  lower panel shows the case
$\beta=10^{-3}$.} \label{fig2}
\end{figure}

\begin{figure}[th]
\includegraphics[width=3.0in,angle=0,clip=true]{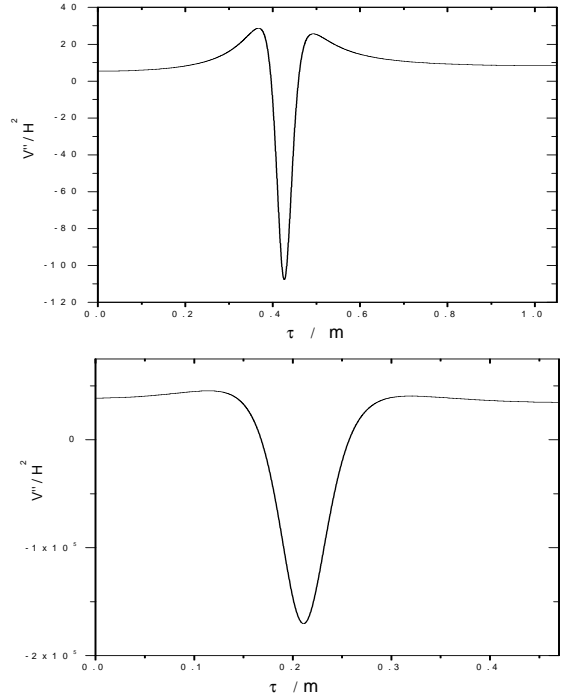}
\vspace{-1.0cm}
\caption{This plot shows how during the tunneling process the inequality $%
|V^{^{\prime\prime}}|\gg H^{2}$ holds. The GR lines represents the
same inequality for Einstein's  Theory of Relativity. The upper
panel shows this case, and the lower panel show the Gauss Bonnet
Brane world case.} \label{fig3}
\end{figure}

We have solved numerically the field equations~(\ref{eqm1}) for
the values gives in table I in the effective Friedmann equation
~(\ref{friedmann2}). The boundary conditions that we have used
were those in which $\phi'=0$ and $a'=1$ at $a=0 $. On the other
hand, we take $m \simeq 10^{-6}$ and the value of the constant
$\delta$ is chosen in such a way that an appropriate amplitude for
density perturbations is obtained. We have also taken the value
$v=3.51$ since they provide the needed $65$ e-folds of inflations
after tunneling has occurred. At $\tau \approx 0$, the scalar
field $\phi =\phi _{T}$ lies in the true vacuum, near the maximum
of the scalar potential $V(\phi )$. At $\tau \neq 0$, the same
field is found close to the false vacuum, but now with a different
value, $\phi =\phi _{F}$. Specifically, the scalar field $\phi $
evolves from some initial value $ \phi _{F}\cong \phi _{i}\approx
3.51$ to the final value $\phi _{T}\cong \phi _{f}$. Numerically
we have found that the CDL instanton does exist, and the Gauss
Bonnet brane world open inflationary universe scenario can be
realized. Table \ref{tab:table1} summarizes our results, which are
compared with those corresponding to Einstein's Theory of
Relativity.
\begin{table}
\caption{\label{tab:table2}This table shows the values of $\phi_F$
and $\phi_T$ for which the CDL instanton  exists. }
\begin{ruledtabular}
\begin{tabular}{lll}
Models & $\phi _{F}$
 & $\phi _{T}$  \\ \hline $\beta=10$ GR
 & $3.51$  & $3.26$  \\ \hline
$\beta=10^{-3}$
GBBW  & $3.51$  & $3.36$  \\
\hline
\end{tabular}
\end{ruledtabular}
\end{table}
Note that the interval of tunneling, specified by $\tau $,
decreases for $\beta=10^{-3}$, but its shapes remain practically
similar. The evolution of the inflaton field as a function of the
Euclidean time is shown in Fig.~\ref{fig2}.
In Fig.~\ref{fig3} we show $|V^{^{\prime \prime }}|\,/\,H^{2}$ as
a function of the Euclidean time $\tau $ for our models. From this
plot we observe that most of the time during the tunneling we
obtain $|V^{^{\prime \prime }}|\,>\,H^{2}$, analogous to that
occurs in  Einstein's Theory of Relativity. Note that, for
$\beta=10$, the peak becomes narrower and deeper, and thus the
above inequality is better satisfied.

\section{\label{sec:level3}Inflation after tunneling}
After the tunneling has occurred, we  make an analytical
continuation to the Lorentzian space-time, and the time evolution
of the scale factor $a(t)$ and the scalar field $\phi (t)$, are
given by
\begin{eqnarray}
\ddot{a} &=&a\beta_{q} m^{2\left( q-1\right) }\left(
\frac{\dot{\phi} ^{2}}{2}+V(\phi )\right) ^{q}\left[
1-\frac{3}{2}q\frac{\dot{\phi }^{2}}{\left( \frac{\dot{\phi}
^{2}}{2}+V(\phi )\right) }\right] ,
\label{eqmL} \\
&&
\begin{array}{ll}
& \ddot{\phi}+3\frac{\dot{a}}{a}\dot{\phi}+V_{,\phi }=0,
\end{array}
\nonumber
\end{eqnarray}
respectively. Here the dots represent derivative with respect to
cosmological time.

Before, to solve this equations we want to discuss the slow-roll
approximation in order to justify in ours models the values of the
parameters $\beta_{q}$'s and the mass of the inflaton field.
During the inflationary epoch the energy density associated to the
scalar field is of the order of the potential, i.e. $\rho\sim V$,
where we have used the slow-roll conditions ($\dot{\phi}^2 \ll
V(\phi)$). Therefore, the Friedmann equation reduces  to
\begin{eqnarray}
H^2\approx\beta_q\,V^q,\label{inf2}
\end{eqnarray}
since we are interested in an inflationary universe model, the
term proportional to $\frac{1}{a^{2}}$ is rapidly diluted. The
field equation of motion for the field $\phi$ becomes
\begin{equation}
3H \dot{\phi}\approx-V_{,\,\phi}. \label{inf3}
\end{equation}

After the tunnel has occurred, the potential energy just
corresponds to the chaotic scalar potential
($V(\phi)\simeq\,m^2\phi^2/2$) by integrating equation
(\ref{inf3}) in the slow roll approximation, we get for arbitrary
$q$
\begin{equation}
\dot{\phi}(t)=-\frac{2^{q/2}\,m^{2-q}}{3\beta_q^{1/2}}\,\phi^{1-q},\label{dotphi}
\end{equation}
and  the full evolution of inflaton field in the Lorentzian era
read as follows
\begin{equation}
\phi(t)=\left[\phi_i^{q}-\frac{2^{q/2}\,\,q\,m^{2-q}}{3\beta_q^{1/2}}\,t\right]^{1/q},
\end{equation}
where $\phi_i$ is a constant and corresponds to the initial value
of the scalar field $\phi$. On the other hand, the scale factor
results to be
\begin{equation}
a(t)=a_i\,\exp\left(\frac{\beta_q^{1/2}\,m^2}{2^{q/2}}\,t
\left[\phi_i^{q}-\frac{2^{q/2}\,\,q\,m^{2-q}}{6\beta_q^{1/2}}\,t\right]\right).
\end{equation}
Where we can see that the energy density associated to the
inflaton field decreases in order to reach the kinetic epoch
($\rho_{kin}\sim a^{-6}$) and the scale factor satisfies the
condition $\ddot{a}>0$.

The consequences of the dynamics of an open inflationary universe,
such as, slow roll parameters, density perturbations and tensor
perturbations are quite complicate to be realized, since it
involves several contributions. As was shown in Refs. \cite{re6,
ramon, delCampo:2004gh, del Campo:2004ee} in an open inflationary
universe model this correction should be somewhat corrected during
the very first stages of the inflationary period, at $N = O(1)$.
For instance, during the first three e-folds the fluctuations of
the inflaton field are not produced inside the bubble. In this way
one may since $N$ should be of order hundred or more in order to
solve the cosmological "puzzles", we may get rid of them and
consider the standard flat-space expressions give correct results
for $N> 3$.

In order to constrain the parameters of our model we considered
the dimensionless slow-roll parameters \cite{Kim:2004gs,
Kim:2004jc, Kim:2004hu, Murray:2006fw},
\begin{equation}
\epsilon=-\frac{\dot{H}}{H^2}=\frac{q}{6\beta_q}\,\frac{V_{\,,\phi}^2}{V^{q+1}},\;\;\;\,
\eta=\frac{V_{\,,\phi\phi}}{3H^2}=\frac{V_{\,,\phi\phi}}{3\beta_q\,V^q}.
\end{equation}
The condition under which we could that ensure that the
inflationary epoch could take place can be summarized with the
inequality $\epsilon<1$ (note that this condition is analogue to
the requirement that $\ddot{a}> 0$). Inflation ends when the field
takes the value,
\begin{equation}
\phi_f\simeq\left[\frac{2^{q+1}\,q\,m^{2(1-q)}}{6\beta_{q}}\right]^{1/2q},
\end{equation}
or equivalently, in term of the value of potential at the end of
inflation,
\begin{equation}
V_f\simeq\,\frac{q\,m^2}{3\,\beta_q}.
\end{equation}
In the following, the subscripts  $*$ and $f$ are used to denote
the epoch when the cosmological scale exits the cosmological
horizon and the end of  inflation, respectively. At the end of
inflation we can compute the number of e-folds which is given by
\begin{equation}
N=\int_{t_*}^{t_f}\,H\,dt'=-3\int_{\phi_{*}}^{\phi_f}\frac{H^2}{V_{,\,\phi}}
d\phi'=-3\,\beta_q\int_{\phi_{*}}^{\phi_f}\frac{V^q}{V_{,\,\phi}}
d\phi',\label{N}
\end{equation}
and using the chaotic potential  the number of e-folds becomes
\begin{equation}
N\simeq\,\frac{3\beta_q}{2q\,m^2}\,\left(\frac{m}{\sqrt{2}}\right)^{2q}\,[
\phi_*^{2q}-\phi_f^{2q}]\simeq\frac{3\beta_q}{2q\,m^2}\,[V_*^q-V_f^{q}]\simeq
\frac{3\beta_q}{2q\,m^2}\,V_*^q-\frac{1}{2}.\label{fold}
\end{equation}
 In order to study the
scalar perturbations of the inflaton fields we considerer that the
conservation of energy-momentum on the brane,
($\nabla^{\nu}T_{\nu\mu}=0$), implies that the adiabatic matter
curvature perturbation $\zeta$ over one uniform density
hypersurface is conserved on large scales, independent of the
gravitational physics \cite{Wand}. Consequently, the amplitudes of
a given mode that re-enters the Hubble radius after inflation is
given by

\begin{equation}
A_S^2\;|_{k=k_*=aH}=\,\left.\frac{H^4}{2\pi^2\dot{\phi}^2}\right|
_{k_*=aH}=\left.\frac{9}{2\pi^2}\,\frac{H^6}{V_{\,,\phi}^2}\right|
_{k_*=aH}=\left.\frac{9\,\beta_q^3}{2\pi^2}
\,\frac{V^{3q}}{V_{\,,\phi}^2}\right|
_{k_*=aH}=\frac{9\,\beta_q^3}{4\pi^2}
\,\frac{V_*^{3q-1}}{m^2}.\label{pert}
\end{equation}

Finally, from Eqs.(\ref{fold}) and (\ref{pert}) we obtain a
constrain on the ratio
\begin{equation}
\frac{m^{2/q-4}}{\beta_q^{1/q}}=\frac{9}{4\pi^2}\,\frac{1}{A_s^2}\,\left[\frac{q\,(2\,N+1)}{3}\right]^{\frac{3q-1}{q}}.\label{constrains}
\end{equation}
This latter expression allows us to fix the values of the
parameters of our model. The needed $N\sim65$ e-folds of
inflations after tunneling has occurred, in order to solve of the
most cosmological puzzles. On the other hand, $A_S \sim 10^{-5}$
in order to fit over the observational data.


Finally, we solve the Lorentzian field Equations (\ref{eqmL})
numerically, we use the following boundary
conditions: $\dot{\phi}(0)=0$, $a(0)=0$ and $\dot{a}%
(0)=1$. As in  \cite{re6},\cite{shiba} and \cite{ramon}, the
scalar field slowly rolls down to its minimum of the effective
potential, and its field starts to oscillate near the minimum.
During this stage, the $N$ e-folds parameter presents different
values for our models. The results are summarized in Figure 4.
\begin{figure}[th]
\includegraphics[width=2.0in,angle=0,clip=true]{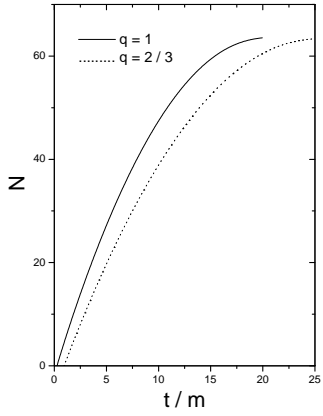}
\vspace{-1.0cm} \caption{The plot shows the number $N$ of e-folds
as a function of the scaled cosmological times $t/m$. The solid
line represents Einstein's Theory of Relativity, and the dashed
line represents the Gauss Bonnet Brane World.} \label{fig4}
\end{figure}


\newpage

\section{conclusion}

In this paper we have studied one-field open inflationary universe
models in which the gravitational effects are described by
Gauss-Bonnet Brane World cosmology. In this kind of theory, the
Friedmann equation have a complicated structure as show by Eq.
(\ref{q3}) and at the limits of Gauss-Bonnet and 4D regime we
obtain an effective Friedmann equation whose form is given by
term: $\beta_{q} \rho^q$, where $q=2/3$ and $q=1$ for the GB and
4D Einstein's Theory of Relativity respectively. We have described
solutions for the Euclidean fields Eqs.(\ref{eqm1}) with an
effective potentials describe in Ref.\cite{re6} in which the CDL
instanton exist. The existence of this instanton becomes
guaranteed since the inequality $|V^{''}|>H^{2}$ is satisfied, and
in the case of $q=2/3$ this inequality is most stronger satisfy
($|V^{''}|\gg H^{2}$) than the Einstein-Hilbert case ($q=1$).
Therefore, the conditions for create an open inflationary
universes are better satisfied in the context of Gauss-Bonnet
Brane world cosmology, and thus, with the slow-roll approximation,
inflationary universes models are realized. For the two values
{\it{i.e.}}($q=1$ and $q=2/3$), $V^{''}$ remains greater than
$H^{2}$ during the first e-folds of inflation.

We also found that the modification of the Friedmann equation
($\rho^q$) improves the characteristics of inflation . This is
accentuated in the number of the e-folds $N$ if we take a little
different values for $\beta_{2/3}$ parameter as it is shown by
Eq.(\ref{fold}).

For $m \sim 10^{-6}$ we have found that, the scalar potential
($V_f$) at the end of inflationary epoch in the Gauss Bonnet Brane
cosmology ($q=2/3$) is greater than the standard general
relativity ($q=1$). In theory with Gauss Bonnet correction the
"velocity" ($\dot{\phi}$) of scalar field decreased in comparison
with the theory without correction term. This remark it is
explicitly show by Eq.(\ref{dotphi}). Consequently, in the Gauss
Bonnet brane world cosmology the time interval required to
generate the appropriate N e-folds numbers of inflations, in order
to solve of the most cosmological puzzles, becomes greater than in
comparison with the general relativity theory ($q=1$). This is
shown in figure 4. Eq. (\ref{constrains}) shows the possible
constraint between the mass parameter of the inflaton field $m$
and the parameter $\beta_q$. This constraint is quite strong
related to observational data through the N e-folds number and the
amplitude of the scalar perturbation ($A_S$). It is well know
that, when $q=1$ the parameter $\beta_1$ is fixed
($\beta_1=\frac{8\pi}{3}\sim O(10)$), then from
Eq.(\ref{constrains}) we can establish the mass parameter of the
inflaton field by $m\sim 10^{-6}$, which corresponds to its
standard value. When $q=2/3$ the parameter $\beta_{2/3}$ is given
by the value of the Gauss Bonnet coupling constant
($\beta_{2/3}=\beta_{2/3}(\alpha)\sim 1/\alpha$). If we consider
the fact that $\alpha \sim 1/g_s$ where $g_s$ is the scale of the
string theory, the value obtained from Eq.(\ref{constrains}) for
the mass parameter is extremely large. Finally, from Eq.
(\ref{constrains}) we can change both parameters in order to fit
ours results with observational data. This changing the constraint
of the parameter $\beta_{2/3}$, and we would also change the value
of the Gauss-Bonnet coupling constant ($\alpha$) arise from string
theory. For example, if we consider the well know value for the
mass of the inflaton field ($m\sim 10^{-6}$), we obtain that
$\beta_{2/3} \sim 10^{-3}$.

\begin{acknowledgments}
S. del C. was supported from COMISION NACIONAL DE CIENCIAS Y
TECNOLOGIA through FONDECYT Grants N$^0$ 1040624, N$^0$  1051086
and N$^0$ 1070306, and was also partially supported by PUCV Grant
N$^0$ 123.787. R. H. was supported by Programa Bicentenario de
Ciencia y Tecnolog\'{\i}a through Grant Inserci\'on de
Investigadores Postdoctorales en la Academia N$^0$ PSD/06. J. S.
was from COMISION NACIONAL DE CIENCIAS Y TECNOLOGIA through
FONDECYT Grant 11060515 and supported by PUCV Grant N$^0$ 123.785.
\end{acknowledgments}

\end{document}